\documentclass[11pt]{article}

 \usepackage{amsmath}
 \usepackage{tabularx}
 \usepackage{array}
 \usepackage{booktabs}
 \usepackage{graphicx}
 \usepackage{longtable}
 \def\CP{{$\cal CP~$}}
 \usepackage{amsfonts}
 \usepackage{float}
\usepackage{hyperref}
\usepackage{bm, geometry, cite, amssymb}
\begin{document}
 
 \title{\LARGE \bf {Connected Sequential Bargmann Invariants and \CP-Sensitive Geometric Correlation Structures in Neutral Meson Systems}}

 \author{
 	Swarup Sangiri \thanks{swarup.phys@gmail.com} \\
   \textit{Physics Department,
	Indian Institute of Technology Kharagpur}\\ \textit{Kharagpur, 721302, India}
}
\date{}
\maketitle
\begin{abstract}
We investigate connected sequential geometric structures in correlated neutral meson systems within the framework of Bargmann invariants. Building upon previously developed third- and fourth-order rephasing-invariant geometric structures involving decay-projected conditional states, we introduce a connected sequential fourth-order Bargmann invariant in which the decay-projected states associated with two decay channels are linked through a direct overlap between the corresponding projected states. This construction incorporates explicit projection-projection correlations within the cyclic overlap chain. The connected sequential invariant encodes the geometric relation between decay-projected states, thereby extending the geometric correlation framework developed for correlated neutral meson systems. To characterize the resulting geometric structures, we define rephasing-invariant ratios that quantify connected sequential correlations and provide a direct comparison with the previously studied disconnected geometric correlations. The behavior of these quantities is analyzed in the regime of small \CP violation using the standard rephasing-invariant interference parameters together with a small-asymmetry expansion. We show that the connected sequential ratios exhibit characteristic scaling behaviors governed by both mixing asymmetry and relative interference-phase alignment, leading to geometric scaling properties distinct from those of the disconnected structures. We further discuss the geometric interpretation of the connected sequential invariants and their possible relevance to correlated neutral meson systems. The resulting framework extends the hierarchy of Bargmann invariant geometric correlations associated with neutral meson mixing and decay, providing a complementary geometric perspective on \CP-sensitive interference phenomena.
\end{abstract}

\newpage

\section{Introduction}

Neutral meson systems provide a particularly rich setting for studying quantum interference phenomena because they naturally combine mixing, oscillation, propagation, and decay dynamics within a single quantum framework \cite{BigiSanda1981, BigiSanda2009}. Systems such as $K^0-\overline{K^0}$, $B^0-\overline{B^0}$, $B_s^0-\overline{B_s^0}$, and $D^0-\overline{D^0}$ exhibit nontrivial interference effects arising from the interplay between mixing and decay amplitudes, making them sensitive probes of phase-dependent correlations and rephasing-invariant structures \cite{Branco1999, Barr2016, Sarkar2008}. Because observable quantities must remain independent of arbitrary phase conventions, neutral meson systems provide a natural arena for investigating geometric and rephasing-invariant aspects of quantum interference.

Alongside the conventional dynamical description of neutral meson mixing and decay, geometric approaches to quantum interference have attracted significant interest \cite{Berry1984, AharonovAnandan1987}. Bargmann invariants and geometric phases provide a natural framework for describing cyclic overlap structures in projective Hilbert space, where physically meaningful quantities depend only on rays rather than on arbitrary phase conventions \cite{Bargmann1964, MukundaSimon1993, Rabei1999}. Since Bargmann invariants (BIs) are intrinsically rephasing invariant, they are particularly well suited for analyzing interference phenomena in neutral meson systems \cite{Mukunda2003}. Within this broader context, the study of \CP violation remains central to understanding the structure of weak interactions and the origin of symmetry-violating phenomena in particle physics \cite{Christenson1964, Wolfenstein1964, Cabibbo1963, KobayashiMaskawa1973}, providing an important physical setting in which geometric and interference-based structures can be explored.

Geometric phases and BIs have previously been investigated in neutral kaon systems, providing a geometric characterization of mixing and \CP-sensitive interference effects \cite{SangiriSarkar2023, Sangiri2024}. Recently, geometric constructions based on BIs have been explored in the context of correlated neutral meson decays \cite{Sangiri2026}. In particular, third- and fourth-order invariants constructed from propagation eigenstates and decay-projected conditional states were shown to encode nontrivial \CP-sensitive interference structures associated with both single-channel and inter-channel decay correlations. These constructions established a geometric interpretation of correlated decay processes in terms of cyclic overlap structures in projective Hilbert space.

Motivated by these developments, it is natural to ask whether additional geometric structures emerge when correlated decay projections are connected sequentially within the cyclic overlap chain itself. Sequential and cascade-like interference phenomena have played an important role in neutral meson physics, especially in processes involving successive propagation and decay stages \cite{Kayser1997, Shen2024}. In conventional cascade \CP analyses, interference structures typically arise through successive mixing and decay stages involving multiple neutral meson sectors and correlated time-dependent evolution.

The present work does not introduce an additional neutral meson propagation stage of this type. Instead, we investigate the geometric consequences of sequentially connected decay projections within a single neutral meson system. The resulting structures are therefore not conventional cascade observables in the usual dynamical sense, but rather connected sequential overlap geometries constructed within the BI framework. This allows the study of how interference information becomes reorganized when projected decay states are directly connected inside cyclic geometric structures.

In this work, we construct a connected sequential fourth-order BI involving direct overlap correlations between projected decay states. Here, the terminology ``connected sequential" reflects the ordered appearance of decay-projected states within the cyclic overlap chain together with the direct overlap relation that links them inside the geometric structure. Building upon the previously studied disconnected fourth-order structures, where different decay channels are correlated through propagation overlaps, the sequential construction introduces explicit projection-projection correlations within the cyclic overlap chain. We further define associated rephasing-invariant geometric ratios that highlight the contribution of these direct projection correlations and provide a quantitative comparison between connected and disconnected geometric structures. These ratios reveal distinct geometric scaling behaviors near the limit of vanishing indirect \CP violation, thereby distinguishing different classes of interference correlations and illustrating how the geometric correlation structure changes when decay-projected states become directly connected within the cyclic overlap chain.

A detailed analysis of the small \CP limit is performed in terms of the standard rephasing-invariant interference parameters together with a systemetic small-asymmetry expansion. In this regime, the sequential geometric ratios exhibit characteristic scaling structures governed not only by mixing asymmetry, but also by the relative interference phases and geometric alignment of the participating decay-projected states. This leads to geometric scaling structures that differ qualitatively from those obtained in the previously studied disconnected constructions. The resulting framework therefore provides a complementary geometric characterization of correlated interference phenomena in neutral meson systems, extending the hierarchy of Bargmann invariant geometric structures through directly connected decay-projected states.

\section{Neutral Meson Mixing and Decay Framework}
Neutral meson systems provide some of the most sensitive laboratories for studying flavor oscillation, quantum interference, and \CP-violating phenomena \cite{Branco1999}. The interplay between mixing and decay in systems such as $K^0$, $D^0$, $B_d^0$, and $B_s^0$ gives rise to observable \CP-violating effects \cite{PDG2024}. The Hilbert space of a generic neutral meson system is spanned by the flavor eigenstates $|P^0\rangle$ and $|\overline{P^0}\rangle$, which carry opposite flavor quantum numbers. While strong and electromagnetic interactions preserve these quantum numbers, weak interactions induce transitions between particles and antiparticles. Consequently, the flavor basis does not diagonalize the effective Hamiltonian responsible for propagation. The physical eigenstates can be conventionally written as 
\begin{subequations}\label{P_HL}
	\begin{align}
	|P_H\rangle &=p|P^0\rangle +q|\overline{P^0}\rangle, \\
	|P_L\rangle &=p|P^0\rangle -q|\overline{P^0}\rangle,
	\end{align}
\end{subequations}
up to arbitrary phase conventions \cite{BigiSanda2009}. Here $|P_H\rangle$ and $|P_L\rangle$ denote the two propagation eigenstates, conventionally labeled heavy and light respectively, and the coefficients $p$ and $q$ characterize the mixing structure of the system and satisfy
\begin{align}
|p|^2+|q|^2=1.
\end{align}
The quantity $\frac{q}{p}$ plays a central role in the description of \CP violation in mixing, also known as indirect \CP violation. In particular, $\left|\frac{q}{p}\right|\ne1$ indicates an asymmetry between particle and antiparticle propagation. In the presence of indirect \CP violation, the propagation eigenstates are generally nonorthogonal, with
\[
\langle P_L|P_H\rangle=|p|^2-|q|^2,
\]
which vanishes only in the \CP-conserving limit in mixing $|q/p|=1$.

In experimental environments such as flavor factories, neutral mesons are frequently produced in correlated pairs through the decay of vector resonances such as $\phi(1020)\rightarrow K^0\overline{K^0}$ and $\Upsilon(4S)\rightarrow B^0\overline{B^0}$ \cite{KLOE2006, BABAR2004, Belle2001}. Because the parent resonance possesses quantum numbers $J^{PC}=1^{--}$, the resulting two-meson state must be antisymmetric under particle exchange. The correlated state therefore takes the entangled form \cite{Lipkin1989, BertlmannGrimus1997, BertlmannHiesmayr2001}
\begin{align}\label{eq:Psi}
|\Psi\rangle=\frac{1}{\sqrt{2}}\left(|P^0\rangle_1|\overline{P^0}\rangle_2-|\overline{P^0}\rangle_1|P^0\rangle_2\right),
\end{align}
where the subscripts 1 and 2 refer to the two mesons. This entangled structure implies that the observation of one meson determines the quantum state of its partner. Hence, correlated decay processes contain interference information that is inaccessible in isolated single-meson decays.

The decay of neutral mesons into a final state $f$ is described by the weak transition operator 
$T$. The corresponding decay amplitudes are defined as \cite{Nir2005} 
\begin{align}\label{eq:decay_amplitude}
A_f=\langle f|T|P^0\rangle, \quad \bar A_f=\langle f|T|\overline{P^0}\rangle.
\end{align}
These amplitudes generally possess both weak and strong phases, and their relative structure governs direct \CP-violating effects in the decay sector. Channels corresponding to flavor tags and approximate \CP eigenstates are particularly important for interference studies in neutral meson phenomenology.

For an entangled meson pair, the decay of one meson induces a projection on the remaining subsystem. If one particle decays into a channel
$f$, the companion meson is projected onto a conditional state determined by the entangled structure and the corresponding decay amplitudes. Performing the projection $\langle f|T|\Psi\rangle$, thereby projecting the first meson onto the decay channel $f$, one obtains the decay-projected state
\begin{align}\label{psi_f}
|\psi_f\rangle=\frac{1}{\sqrt{2}}\left(A_f|\overline{P^0}\rangle-\bar A_f|P^0\rangle\right),
\end{align}
up to an overall normalization factor. The overall factor $\frac{1}{\sqrt{2}}$ arises directly from the normalization of the entangled state in Eq.~(\ref{eq:Psi}) and does not imply that the conditional state itself is normalized. Since only relative overlap structures and phases are relevant for the geometric quantities considered below, the overall normalization of the conditional state will not play a role in the subsequent analysis. In Eq.~(\ref{psi_f}), the subsystem labels are no longer written explicitly because, after the projection, the remaining state refers entirely to the surviving meson. The structure of $|\psi_f\rangle$ therefore encodes the correlation inherited from the original entangled state together with the interference information contained in the decay amplitudes. 

Similarly, for another decay channel 
$g$,
\begin{align}\label{psi_g}
|\psi_g\rangle=\frac{1}{\sqrt{2}}\left(A_g|\overline{P^0}\rangle-\bar A_g|P^0\rangle\right).
\end{align}
These decay-projected states play a central role in the geometric structures considered later in this work. In particular, they provide the intermediate states entering the construction of higher-order BIs associated with correlated and sequential decay processes. The overlap structure between such projected states naturally encodes interference between mixing and decay amplitudes and forms the basis for the geometric correlation framework developed in the following sections.

\section{Geometric Correlation Structures and Bargmann Invariants in Neutral Meson Systems}
Bargmann invariants (BIs) and geometric phases provide a natural framework for describing rephasing-invariant structures in quantum systems. In projective Hilbert space, BIs characterize cyclic sequences of state overlaps and encode geometric information associated with interference phenomena \cite{Bargmann1964, MukundaSimon1993}. For a sequence of quantum states ${|\psi_1\rangle,|\psi_2\rangle,\dots,|\psi_n\rangle}$, the corresponding $n$th-order BI is defined as
\begin{align}
\Delta_n=(\psi_1,\psi_2)(\psi_2,\psi_3)\cdots(\psi_n,\psi_1),
\end{align}
where $(\psi_i,\psi_j)\equiv\langle\psi_i|\psi_j\rangle$ and no two consecutive states are orthogonal. These quantities are invariant under independent phase redefinitions of the participating states and therefore depend only on the associated rays in projective Hilbert space. The phase of the invariant corresponds to a geometric phase accumulated along the closed cyclic sequence of states. 

In this section, we briefly summarize the geometric framework and the BI structures developed previously in Ref.~\cite{Sangiri2026}, which form the basis for the connected sequential constructions introduced later. These structures established a geometric description of single-channel and inter-channel decay correlations in terms of cyclic overlap sequences involving propagation eigenstates and decay-projected states. Within the neutral meson framework introduced in the previous section, the decay-projected conditional states $|\psi_f\rangle$ and $|\psi_g\rangle$ allow the construction of nontrivial geometric cycles involving propagation and decay processes. The simplest nontrivial structure is obtained from the cyclic sequence $P_H\rightarrow\psi_f\rightarrow P_L\rightarrow P_H$, which defines the third-order BI
\begin{align}\label{BI3}
\Delta_3=(P_H,\psi_f)(\psi_f,P_L)(P_L,P_H).
\end{align}
Using the explicit forms of the propagation eigenstates (Eq.~(\ref{P_HL}))  and the decay-projected state (Eq.~(\ref{psi_f})), the invariant can be written as
\begin{subequations}\label{eq:BI_3}
 \begin{align}
\Delta_3 &=\frac12 (|p|^2-|q|^2) (q^*A_f - p^*\bar A_f) (-qA_f^* - p\bar A_f^*)\\
&= \frac12(|p|^2-|q|^2) \Big[ |p|^2|\bar A_f|^2 - |q|^2|A_f|^2 + 2i\,\mathrm{Im}(p^*q\,\bar A_f A_f^*) \Big].
\end{align}   
\end{subequations}

The structure of $\Delta_3$ directly links the geometric invariant to the mixing parameters and decay amplitudes of the neutral meson system. In particular, the imaginary contribution $\mathrm{Im}(p^*q\,\bar A_f A_f^*)$ arises from interference between mixing and decay phases and therefore carries \CP-sensitive information. The appearance of the factor $(P_L,P_H)=|p|^2-|q|^2$ shows that the third-order invariant is directly sensitive to the nonorthogonality of the propagation eigenstates and vanishes in the exact \CP-conserving limit, where propagation eigenstates become orthogonal. The associated geometric phase, given by $\arg(\Delta_3)$, corresponds to the phase accumulated along the closed triangular path in ray space.

A higher-order geometric structure is obtained by incorporating two decay channels simultaneously. Considering the cyclic sequence $P_H\rightarrow\psi_f\rightarrow P_L\rightarrow \psi_g\rightarrow P_H$, one obtains the fourth-order BI
\begin{align}\label{eq:Delta4}
\Delta_4 = (P_H,\psi_f) (\psi_f,P_L) (P_L,\psi_g) (\psi_g,P_H).
\end{align}
Evaluating the overlaps yields
\begin{subequations}\label{eq:BI_4}
 \begin{align}
\Delta_4 &=\frac{1}{4}\left[ (q^*A_f - p^*\bar A_f) (-qA_f^* - p\bar A_f^*) (-p^*\bar A_g - q^*A_g) (qA_g^* - p\bar A_g^*)\right]\\
&= \frac{1}{4} \left[a_f a_g + b_f b_g + i(a_g b_f - a_f b_g)\right],
\end{align}   
\end{subequations}
where
\begin{align}
a_i = |p|^2|\bar A_i|^2 - |q|^2|A_i|^2, \qquad b_i = 2\mathrm{Im}(p^*q\bar A_i A_i^*), \quad (i=f,g).
\end{align}

Unlike the third-order invariant, the fourth-order structure correlates the \CP-sensitive properties of two distinct decay channels. The appearance of mixed combinations such as $a_gb_f-a_fb_g$ demonstrates that $\Delta_4$ encodes nontrivial inter-channel interference effects that cannot be directly inferred from the corresponding single-channel geometric structures alone.
To isolate these correlated geometric structures, a ratio was introduced \cite{Sangiri2026},
\begin{align}\label{eq:R}
R = \frac{\Delta_4}{\Delta_3(f)\Delta_3(g)}.
\end{align}
This construction factors out the leading single-channel geometric contributions and emphasizes the correlated interference structure shared between the two decay channels. Expressed in terms of the quantities $a_i$ and $b_i$, the ratio takes the form
\begin{align}
R =\frac{a_f a_g + b_f b_g + i(a_g b_f - a_f b_g)}{(|p|^2 - |q|^2)^2 \left[a_f a_g - b_f b_g + i(a_f b_g + a_g b_f)\right]}.
\end{align}
An important feature of this ratio is the enhancement factor
\[
\frac{1}{(|p|^2-|q|^2)^2},
\]
which becomes large near the \CP-conserving limit $|p|^2\simeq|q|^2$. Consequently, the ratio can exhibit enhanced sensitivity to small deviations from \CP symmetry in the mixing sector. Furthermore, the symmetric and antisymmetric combinations appearing in the numerator and denominator encode nontrivial correlations between the interference structures of different decay channels. In the exact limit of vanishing indirect \CP violation ($|q/p|=1$), the ratio becomes ill-defined due to the vanishing of the third-order invariants, although in the absence of \CP violation such a \CP-sensitive probe is not required.

These results establish a geometric hierarchy of rephasing-invariant structures associated with neutral meson mixing and decay. The third-order invariant probes single-channel interference geometry, while the fourth-order invariant and the ratio $R$ capture correlated multi-channel structures. This framework provides the basis for extending the analysis to sequential geometric structures generated by successive decay projections, where additional overlap correlations give rise to more intricate geometric organization of \CP-sensitive interference phenomena.

\section{Sequential Decay Geometry and Correlated Projection Structures}
Sequential decay processes and cascade-decay interference phenomena have played an important role in the study of neutral meson systems \cite{Kayser1997, Shen2024, BigiSanda2009}. In conventional cascade \CP analyses, interference effects arise through successive propagation and decay stages involving multiple neutral meson sectors. Typical examples include processes such as $B^0 \rightarrow J/\psi K^0$ \cite{BABAR2004, Belle2001}, followed by the subsequent mixing and decay of the neutral kaon sector. In such systems, the interference structure receives contributions from multiple propagation stages, leading to correlated time-dependent \CP-sensitive effects associated with both mixing and decay dynamics.

The present work does not introduce an additional neutral meson propagation sector of this type. Instead, we focus on the geometric structures generated by successive decay projections within a single neutral meson system. Although the dynamical structure differs from conventional cascade-decay formalisms, the resulting overlap chains still possess a sequential character. In particular, the projected decay states enter the cyclic overlap structures in an ordered manner, generating correlated geometric relations among decay-projected states within projective Hilbert space. It is this ordered appearance of decay-projected states that motivates the use of the term ``sequential" in the present work. In the constructions developed later, successive projected states are linked through an explicit overlap, introducing a direct geometric connection within the cyclic overlap chain. The terminology ``connected sequential" is therefore used to emphasize both the ordered organization of decay projections and the direct overlap relation that links them inside the geometric structure. These terms refer to the organization of the cyclic overlap geometry and should not be interpreted as implying an additional dynamical cascade-decay stage or a second neutral-meson propagation sector.

From the perspective of BIs, sequential overlap structures provide a natural framework for studying how interference information is encoded across correlated decay channels. The previously discussed third- and fourth-order invariants already encode nontrivial geometric correlations between propagation eigenstates and decay-projected states. In particular, the fourth-order structure
\[
\Delta_4
=
(P_H,\psi_f)
(\psi_f,P_L)
(P_L,\psi_g)
(\psi_g,P_H)
\]
correlates two decay channels through cyclic heavy-light transition overlaps. This construction characterizes an inter-channel geometric structure in which the decay projections participate through separate overlaps involving the heavy and light propagation eigenstates \cite{Sangiri2026}.

Motivated by the sequential organization of these correlated overlap structures, it is natural to consider whether additional geometric structures arise when the projected decay states themselves become directly connected inside the cyclic overlap chain. Such a construction introduces a distinct organization of the interference geometry, since the decay projections are no longer correlated only through propagation overlaps, but also through their mutual geometric relation in projective Hilbert space.

The sequential structures introduced in the following sections are motivated by this observation. Rather than extending the conventional cascade-decay formalism, the present framework explores the geometric consequences of connected sequential overlap structures generated by correlated decay projections. This provides a complementary geometric viewpoint in which inter-channel interference is encoded through connected projection correlations in addition to the propagation-induced overlap structure.

\section{Connected Sequential Geometric Structures and Bargmann Invariants in Neutral Meson Systems}
The geometric structures discussed in the previous sections were constructed from cyclic overlap sequences involving propagation eigenstates and decay-projected conditional states in correlated neutral meson systems. We now extend this framework to a connected sequential geometric structure in which two decay-projected states are linked directly through an overlap factor $(\psi_f,\psi_g)$.
We consider the sequential cyclic structure $P_H\rightarrow\psi_f\rightarrow \psi_g\rightarrow P_L\rightarrow P_H$, which defines a connected loop in the projective Hilbert space of the neutral meson system. Unlike the previously constructed fourth-order invariant, where the decay channels $f$ and $g$ entered through separate overlap structures involving the heavy and light propagation eigenstates, the present construction directly correlates the projected states $|\psi_f\rangle$ and $|\psi_g\rangle$. The resulting invariant therefore probes explicitly connected geometric correlations between the decay-projected states associated with different decay channels. 

\begin{figure}[H]
\centering
\includegraphics[width=0.55\textwidth]{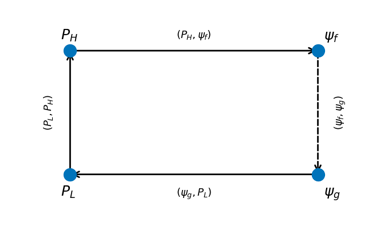}
\caption{
Schematic representation of the connected sequential fourth-order BI
$\Delta_4^{seq}
=(P_H,\psi_f)(\psi_f,\psi_g)(\psi_g,P_L)(P_L,P_H)$.
The cyclic sequence
$P_H\rightarrow\psi_f\rightarrow\psi_g\rightarrow P_L\rightarrow P_H$
defines the corresponding closed overlap path in projective Hilbert space. The dashed segment highlights the direct overlap $(\psi_f,\psi_g)$ between the decay-projected states that distinguishes the connected sequential structure from the previously studied disconnected fourth-order invariant.
}
\label{fig:seq_cycle}
\end{figure}

The corresponding fourth-order BI is given by
\begin{align}
\Delta^{seq}_4=(P_H,\psi_f)(\psi_f,\psi_g)(\psi_g,P_L)(P_L,P_H).
\end{align}
The geometric structure associated with the connected sequential fourth-order BI is illustrated schematically in Fig.~\ref{fig:seq_cycle}.
Using the propagation eigenstates in Eq.~(\ref{P_HL}), together with the decay-projected states in Eq.~(\ref{psi_f}) and Eq.~(\ref{psi_g}), the relevant overlaps become
\begin{subequations}
	\begin{align}
		(P_H,\psi_f) &=\frac{1}{\sqrt{2}}(q^*A_f-p^*\bar A_f),\\
        (\psi_f,\psi_g) &=\frac{1}{2}(A_f^*A_g+\bar A_f^*\bar A_g), \\
		(\psi_g,P_L) &=\frac{1}{\sqrt{2}}(-qA_g^*-p\bar A_g^*),\\
		(P_L,P_H) &=|p|^2-|q|^2.
	\end{align}
\end{subequations}
Substituting these overlaps into the cyclic product yields
\begin{align}\label{eq:BI_seq}
\Delta_4^{seq}=\frac{1}{4}(|p|^2-|q|^2)(q^*A_f-p^*\bar A_f)(-qA_g^*-p\bar A_g^*)(A_f^*A_g+\bar A_f^*\bar A_g).
\end{align}
The structure of this invariant differs qualitatively from the previously obtained fourth-order invariant. In particular, the factor
$(A_f^*A_g+\bar A_f^*\bar A_g)$ directly correlates the decay channels through the overlap between the projected states themselves. The invariant therefore incorporates mixing-induced \CP-sensitive effects together with direct geometric overlap correlations between decay-projected states. As in the previous constructions, the invariant depends only on cyclic overlaps between rays and therefore remains invariant under arbitrary independent phase redefinitions of the participating states. Since $\Delta_4^{seq}$ is proportional to the overlap factor $(P_L,P_H)=|p|^2-|q|^2$, the invariant vanishes in the exact indirect \CP-conserving limit $|q/p|=1$.

Having constructed the connected sequential invariant, it is natural to compare it with the geometric correlation structures introduced previously. In the earlier framework, the ratio $R$ (Eq.~(\ref{eq:R})) was introduced as a measure of correlated \CP-sensitive geometric structures between two decay channels \cite{Sangiri2026}. We now define the analogous quantity for the connected projection geometry,
\begin{align}
    R_{seq}=\frac{\Delta_4^{seq}}{\Delta_3(f)\Delta_3(g)}.
\end{align}
Using Eq.~(\ref{eq:BI_3}), and Eq.~(\ref{eq:BI_seq}), the ratio takes the form
\begin{align}\label{eq_R_seq}
R_{seq}=\frac{(A_f^*A_g+\bar A_f^*\bar A_g)}{(|p|^2-|q|^2)(-qA_f^* - p\bar A_f^*)(q^*A_g-p^*\bar A_g)}.
\end{align}
The ratio $R_{seq}$ provides a rephasing-invariant measure of the connected sequential geometric correlations between the decay projections associated with the channels $f$ and $g$. Unlike the previously constructed ratio
$R$, which exhibited a quadratic enhancement proportional to $\frac{1}{(|p|^2-|q|^2)^2}$, the connected sequential ratio contains only a linear enhancement
\[
    R_{seq}\sim\frac{1}{(|p|^2-|q|^2)}.
\]
This reduction originates from the direct sequential overlap correlation between the projected states $|\psi_f\rangle$ and $|\psi_g\rangle$, which leads to the cancellation of one power of the mixing-induced suppression factor appearing in the disconnected geometric structure. Unlike the disconnected construction, the connected sequential structure retains explicit information about the relative geometric orientation of the decay-projected states through the overlap $(\psi_f,\psi_g)$. Consequently, $R_{seq}$ can exhibit enhanced sensitivity to small deviations from the \CP-conserving limit while encoding a geometrically distinct interference structure associated with sequential projection correlations. Just as in the previously constructed ratio $R$, the quantity $R_{seq}$ also becomes ill-defined in the exact \CP-conserving limit $|q/p|=1$, where the overlap between the heavy and light propagation eigenstates vanishes. This is not pathological, but rather reflects the disappearance of the underlying mixing-induced interference structure required for defining the geometric ratio. Consequently, $R_{seq}$ is meaningful only away from the exact \CP-symmetric limit, where nontrivial geometric correlations persist. 

It is also useful to compare the connected sequential invariant with the previously constructed fourth-order invariant in Eq.~(\ref{eq:BI_4}). We therefore introduce the ratio
\begin{align}
    \widetilde R_{seq}=\frac{\Delta_4^{seq}}{\Delta_4},
\end{align}
which provides a direct comparison between connected sequential geometric correlations and the earlier disconnected correlation structure. 

The ratio $\widetilde R_{seq}$ isolates the effect of the additional sequential overlap correlation $(\psi_f,\psi_g)$ and therefore measures the deviation between disconnected and explicitly connected geometric structures in the projective Hilbert space of neutral mesons. The detailed properties and comparative behavior of these geometric correlation measures are analyzed in the following section.

\section{Comparative Structure of Sequential and Disconnected Invariants}
Having constructed the connected sequential fourth-order invariant and introduced the ratio $\widetilde R_{seq}$ in the previous section, we now examine its explicit structure and compare it with the previously derived geometric ratio $R$ defined in Eq.~(\ref{eq:R}). The two quantities probe different classes of geometric correlations in the neutral meson system and exhibit distinct scaling sensitive to \CP-violating mixing effects.

Using the expressions obtained for the sequential and disconnected fourth-order invariants in Eq.~(\ref{eq:BI_4}) and Eq.~(\ref{eq:BI_seq}), the common overlap factor $ (q^*A_f-p^*\bar A_f)$
cancels in the ratio, yielding
\begin{align}\label{eq:R_tilde}
\widetilde R_{seq}= \frac{
(|p|^2-|q|^2)
(A_f^*A_g+\bar A_f^*\bar A_g)
(-qA_g^*-p\bar A_g^*)
}{
(-qA_f^*-p\bar A_f^*)
(-p^*\bar A_g-q^*A_g)
(qA_g^*-p\bar A_g^*)
}.
\end{align}
The resulting expression shows explicitly that the sequential geometric structure introduces a direct overlap correlation between the projected decay states through the combination $(A_f^*A_g+\bar A_f^*\bar A_g)$. Unlike the disconnected fourth-order invariant, where the two decay channels are connected only indirectly through the propagation eigenstates, the sequential construction correlates the projected states themselves within the cyclic overlap structure. Consequently, $\widetilde R_{seq}$ provides a comparative characterization of the geometric correlation structure induced by this direct sequential connection.

An important distinction between the two ratios is their dependence on the mixing asymmetry factor $(|p|^2-|q|^2)$. The previously constructed ratio $R$ exhibited the enhanced behavior
\begin{align}
R \sim \frac{1}{(|p|^2-|q|^2)^2},
\end{align}
which originated from the appearance of two independent third-order invariants in the denominator, each contributing a factor of $(|p|^2-|q|^2)$. By contrast, for finite decay-amplitude overlap factors, the sequential ratio scales as
\begin{align}
\widetilde R_{seq} \sim (|p|^2-|q|^2),
\end{align}
so that the quadratic inverse enhancement disappears entirely. This demonstrates that the connected and disconnected geometric structures belong to distinct scaling classes near the \CP-conserving limit.

This structural difference has a clear geometric origin. In the disconnected construction, the two projected decay channels remain separated inside the cyclic overlap structure and are connected only through the heavy-light propagation overlaps, producing the enhanced sensitivity near the \CP-conserving limit. In the sequential construction, however, the explicit overlap between the projected states $|\psi_f\rangle$ and $|\psi_g\rangle$ introduces a direct geometric connection between the projected decay channels. The resulting ratio therefore probes a different aspect of the projective-space geometry associated with correlated neutral meson decays.

The behavior of $\widetilde R_{seq}$ in the limit of vanishing indirect \CP violation is also qualitatively different from that of $R$. Since $\Delta_4^{seq}\propto (|p|^2-|q|^2)$, the sequential invariant vanishes when $|p|=|q|$. Accordingly, $\widetilde R_{seq}\to 0$ in the exact \CP-symmetric limit. This reflects the fact that the connected sequential geometric structure becomes trivial once the heavy-light overlap vanishes. The sequential invariant therefore constitutes a \CP-sensitive geometric structure whose existence depends directly on the mixing asymmetry of the neutral meson system.

\section{Small \CP Expansion and Geometric Scaling Structure}
Having constructed the connected sequential invariant and the corresponding geometric ratios, we next analyze their behavior in the small \CP regime using a rephasing-invariant parametrization of mixing and decay interference. This allows both the geometric phase structure and the characteristic scaling behavior of the sequential correlations to be examined systematically. The geometric phase associated with the sequential invariant of Eq.~(\ref{eq:BI_seq}) is given by
\begin{align}
\gamma_{seq}=\arg\left(\Delta_4^{seq}\right)
\end{align}
corresponding to the geometric phase accumulated along the closed sequence $P_H\rightarrow\psi_f\rightarrow \psi_g\rightarrow P_L\rightarrow P_H$. Using the explicit form of the invariant, the phase may be written as
\begin{align}
    \gamma_{seq}=\arg\left[(q^*A_f-p^*\bar A_f)(qA_g^*+p\bar A_g^*)(A_f^*A_g+\bar A_f^*\bar A_g)\right],
\end{align}
since the overall factor $-\frac{1}{4}(|p|^2-|q|^2)$ is real and therefore does not contribute to the geometric phase modulo $\pi$.

The behavior of both the invariant and the corresponding geometric phase becomes particularly transparent in the regime of small \CP violation. We consider a regime in which both indirect and direct \CP violation are small, so that mixing asymmetry is small and the decay amplitude magnitudes are approximately equal: $|p|\simeq |q|$ and $|A_i|\simeq|\bar A_i|$. In this limit, the quantity $|p|^2-|q|^2$ becomes parametrically small, implying that the magnitude of the invariant is suppressed according to
\begin{align}
    |\Delta_4^{seq}|\sim(|p|^2-|q|^2),
\end{align}
assuming the remaining overlap factors do not vanish. Consequently, the invariant vanishes in the exact \CP-conserving limit $|q/p|=1$, where the overlap between the heavy and light propagation eigenstates vanishes: $(P_L,P_H)=|p|^2-|q|^2=0$. This shows that the sequential invariant is directly tied to mixing-induced \CP asymmetry. Geometrically, the cyclic overlap chain degenerates once the heavy and light eigenstates become orthogonal in the exact \CP-symmetric limit.

However, although the magnitude of the invariant becomes suppressed, the phase structure remains nontrivial. Under the condition $|A_i|\simeq|\bar A_i|$, to leading order in magnitude asymmetry, the decay amplitudes may still contain relative interference phases. Writing schematically
\begin{align}
    \bar A_i\simeq A_ie^{i\phi_i},
\end{align}
the overlap structures become
\begin{align}
    q^*A_f-p^*\bar A_f\sim A_f(q^*-p^*e^{i\phi_f}),
\end{align}
and
\begin{align}
    A_f^*A_g+\bar A_f^*\bar A_g\sim A_f^*A_g(1+e^{i(\phi_g-\phi_f)}).
\end{align}
These combinations remain generically nonvanishing even when the magnitudes of the decay amplitudes are approximately equal. Therefore, the geometric phase $\gamma_{seq}$ continues to encode nontrivial interference information associated with the relative phases of the decay channels and the mixing sector. This reveals an important feature of the sequential geometric structure. The suppression of the invariant in the small \CP regime is controlled primarily by the heavy-light overlap factor, whereas the phase remains governed by the relative orientation of the projected decay states in the underlying projective Hilbert space.

An important feature of the sequential construction is the appearance of the overlap term $(A_f^*A_g+\bar A_f^*\bar A_g)$, which directly correlates the projected states associated with the two decay channels. The sequential invariant therefore probes not only mixing-induced interference effects but also the direct geometric relation between successive decay projections. In this sense, the sequential cyclic structure retains nontrivial phase-sensitive geometric information even when both direct and indirect \CP violation are small.

To further investigate the behavior of the sequential geometric structure near the \CP-conserving regime, it is useful to analyze the ratio $R_{seq}$ derived in Eq.~(\ref{eq_R_seq}) in the limit of small \CP violation. For this purpose, it is convenient to express the ratio in terms of the standard rephasing-invariant quantities \cite{Nir2015}
\begin{align}
\lambda_i=\frac{q}{p}\frac{\bar A_i}{A_i},
\end{align}
which encode the combined effects of mixing and decay interference. Using these definitions, the decay amplitudes may be rewritten as
\begin{align}
\bar A_i=\frac{p}{q}\lambda_i A_i,
\end{align}
up to overall phase conventions which do not affect the rephasing-invariant structure of the ratio.

Substituting these expressions into Eq.~(\ref{eq_R_seq}), and canceling the overall decay-amplitude normalization factors, the sequential ratio may be written in the form
\begin{align}
R_{seq}
=
-
\frac{
|q|^2+|p|^2\lambda_f^*\lambda_g
}{
(|p|^2-|q|^2)
(|q|^2+|p|^2\lambda_f^*)
(|q|^2-|p|^2\lambda_g)
}.
\end{align}
To discuss the scaling behavior near the \CP-conserving limit, we introduce the parametrization
\begin{align}
|p|^2=\frac12(1+\epsilon_m),
\qquad
|q|^2=\frac12(1-\epsilon_m),
\qquad
|\epsilon_m|\ll1,
\end{align}
which provides a convenient way to characterize small deviations from the \CP-conserving limit  $|p|^2=|q|^2=\frac12$. Here $\epsilon_m=|p|^2-|q|^2$ measures the mixing-induced deviation from exact \CP symmetry.

For the interference quantities, we parametrize
\begin{align}
\lambda_i=(1+\kappa_i)e^{i\phi_i},
\qquad
|\kappa_i|\ll1,
\end{align}
where $\kappa_i$ are taken to be real small parameters characterizing deviations of $|\lambda_i|$ from unity and $\phi_i$ denotes the corresponding interference phase. No hierarchy is assumed between the small quantities $\epsilon_m$, $\kappa_f$ and $\kappa_g$.

Substituting these expressions into the numerator of $R_{seq}$ gives
\begin{align}
N
&=
|q|^2+|p|^2\lambda_f^*\lambda_g \nonumber\\
&=
\frac12(1-\epsilon_m)
+
\frac12(1+\epsilon_m)
(1+\kappa_f)(1+\kappa_g)
e^{i(\phi_g-\phi_f)}.
\end{align}
Retaining only leading-order contributions and neglecting terms of order $\mathcal O(\epsilon_m^2,\kappa_i^2,\epsilon_m\kappa_i)$, one obtains
\begin{align}
N
\simeq
\frac12
\left[
1-\epsilon_m
+
(1+\epsilon_m+\kappa_f+\kappa_g)e^{i(\phi_g-\phi_f)}
\right].
\end{align}

Similarly, the denominator factors reduce to
\begin{align}
|q|^2+|p|^2\lambda_f^*
&\simeq
\frac12
\left[
1-\epsilon_m
+
(1+\epsilon_m+\kappa_f)e^{-i\phi_f}
\right], \\
|q|^2-|p|^2\lambda_g
&\simeq
\frac12
\left[
1-\epsilon_m
-
(1+\epsilon_m+\kappa_g)e^{i\phi_g}
\right],
\end{align}
where higher-order contributions of order $\mathcal O(\epsilon_m^2,\kappa_i^2,\epsilon_m\kappa_i)$ have again been neglected.

The sequential ratio therefore becomes
\begin{align}
R_{seq}
\simeq
-2
\frac{
1-\epsilon_m
+
(1+\epsilon_m+\kappa_f+\kappa_g)e^{i(\phi_g-\phi_f)}
}{
\epsilon_m
\left[
1-\epsilon_m
+
(1+\epsilon_m+\kappa_f)e^{-i\phi_f}
\right]
\left[
1-\epsilon_m
-
(1+\epsilon_m+\kappa_g)e^{i\phi_g}
\right]
}.
\end{align}

Several important features follow directly from this structure. First, the ratio exhibits an overall enhancement proportional to
\begin{align}
R_{seq}
\sim
\frac1{\epsilon_m},
\end{align}
provided the remaining overlap factors remain nonvanishing. This indicates that the sequential geometric structure becomes increasingly sensitive near the \CP-conserving limit. This enhancement originates from the heavy-light overlap factor $(P_L,P_H)=|p|^2-|q|^2$, and is weaker than the quadratic enhancement $R\sim\frac{1}{\epsilon_m^2}$ obtained for the disconnected geometric ratio \cite{Sangiri2026}.

Furthermore, unlike disconnected geometric constructions, the sequential ratio depends explicitly on the relative phase difference $(\phi_g-\phi_f)$, through the numerator structure. The quantity $R_{seq}$ therefore probes the relative geometric alignment between the projected decay channels themselves. Additional enhancement may occur when the interference phases approach alignment, so that overlap factors of the form
$1-e^{i\phi_i}$ become small. In this regime, the denominator overlap structures become further suppressed, indicating that the sequential geometric correlation is simultaneously sensitive to mixing asymmetry and to the relative interference phases associated with the decay projections. 

To further compare the connected sequential geometric structure with the previously constructed disconnected fourth-order invariant, we now analyze the ratio $\widetilde R_{seq}$ introduced in Eq.~(\ref{eq:R_tilde}) in the regime of small \CP violation. Rewriting Eq.~(\ref{eq:R_tilde}) in terms of the rephasing-invariant quantities $\lambda_i$, and removing overall phase factors that cancel in the ratio, one obtains
\begin{align}
\widetilde R_{seq}
=
-
(|p|^2-|q|^2)\,
|q|^2
\frac{
\left[
1+\frac{|p|^2}{|q|^2}\lambda_f^*\lambda_g
\right]
\left(
|q|^2+|p|^2\lambda_g^*
\right)
}{
\left(
|q|^2+|p|^2\lambda_f^*
\right)
\left(
|q|^2+|p|^2\lambda_g
\right)
\left(
|q|^2-|p|^2\lambda_g^*
\right)
}.
\end{align}

Using the parametrizations introduced previously and retaining only terms linear in the small quantities $\epsilon_m$, $\kappa_f$ and $\kappa_g$, while neglecting higher-order contributions of order $\mathcal O(\epsilon_m^2,\kappa_i^2,\epsilon_m\kappa_i)$, the numerator factor becomes
\begin{align}
1+\frac{|p|^2}{|q|^2}\lambda_f^*\lambda_g
\simeq
1+
(1+2\epsilon_m+\kappa_f+\kappa_g)
e^{i(\phi_g-\phi_f)},
\end{align}
while the remaining overlap structures reduce to
\begin{align}
|q|^2+|p|^2\lambda_i
&\simeq
\frac12
\left[
1-\epsilon_m
+
(1+\epsilon_m+\kappa_i)e^{i\phi_i}
\right],\\
|q|^2+|p|^2\lambda_i^*
&\simeq
\frac12
\left[
1-\epsilon_m
+
(1+\epsilon_m+\kappa_i)e^{-i\phi_i}
\right],\\
|q|^2-|p|^2\lambda_g^*
&\simeq
\frac12
\left[
1-\epsilon_m
-
(1+\epsilon_m+\kappa_g)e^{-i\phi_g}
\right].
\end{align}

Substituting these expressions into $\widetilde R_{seq}$ yields
\begin{subequations}
   \begin{align}
\widetilde R_{seq}
&\simeq
\frac{ -
2\epsilon_m(1-\epsilon_m)
\left[
1+
(1+2\epsilon_m+\kappa_f+\kappa_g)e^{i(\phi_g-\phi_f)}
\right]
\left[
1-\epsilon_m
+
(1+\epsilon_m+\kappa_g)e^{-i\phi_g}
\right]
}{
\left[
1-\epsilon_m
+
(1+\epsilon_m+\kappa_f)e^{-i\phi_f}
\right]
\left[
1-\epsilon_m
+
(1+\epsilon_m+\kappa_g)e^{i\phi_g}
\right]
\left[
1-\epsilon_m
-
(1+\epsilon_m+\kappa_g)e^{-i\phi_g}
\right]
}.\\
&\simeq
\frac{-
2\epsilon_m
\left[
1+
(1+2\epsilon_m+\kappa_f+\kappa_g)e^{i(\phi_g-\phi_f)}
\right]
\left[
1-\epsilon_m
+
(1+\epsilon_m+\kappa_g)e^{-i\phi_g}
\right]
}{
\left[
1-\epsilon_m
+
(1+\epsilon_m+\kappa_f)e^{-i\phi_f}
\right]
\left[
1-\epsilon_m
+
(1+\epsilon_m+\kappa_g)e^{i\phi_g}
\right]
\left[
1-\epsilon_m
-
(1+\epsilon_m+\kappa_g)e^{-i\phi_g}
\right]
}.
\end{align} 
\end{subequations}

Several important features follow from this structure. Unlike the sequential ratio $R_{seq}$, which exhibits an enhancement proportional to $1/\epsilon_m$, the ratio $\widetilde R_{seq}$ contains an explicit overall suppression proportional to
\begin{align}
\widetilde R_{seq} \sim \epsilon_m,
\end{align}
for finite overlap factors. This behavior originates from the fact that both the connected sequential invariant and the disconnected fourth-order invariant already contain the heavy-light overlap factor $(P_L,P_H)=|p|^2-|q|^2$, so that the singular enhancement present in $R_{seq}$ no longer appears. At the same time, the denominator contains the overlap structure $|q|^2-|p|^2\lambda_g^*$, which becomes suppressed near the phase-alignment limit $\lambda_g \to 1$. Consequently, the behavior of $\widetilde R_{seq}$ is governed by a competition between mixing-induced \CP asymmetry and interference-phase alignment. In particular, while $\widetilde R_{seq}$ is generically suppressed near exact \CP symmetry, this suppression may be weakened when the projected decay-channel phases become geometrically aligned.

The ratio $\widetilde R_{seq}$ therefore probes the relative strength of connected sequential geometric correlations compared to disconnected fourth-order structures, retaining sensitivity to both mixing asymmetry and relative decay-channel coherence in the small \CP regime.

\section{Geometric Interpretation of Connected Sequential \CP Structures}
The sequential invariant introduced in the previous section provides a complementary geometric organization of \CP-sensitive interference structures in neutral meson systems. The previously constructed fourth-order invariant in Eq.~(\ref{eq:Delta4}),
\[
\Delta_4=(P_H,\psi_f)(\psi_f,P_L)(P_L,\psi_g)(\psi_g,P_H),
\]
and the sequential invariant introduced in Eq.~(\ref{eq:BI_seq}),
\[
\Delta_4^{seq}
=(P_H,\psi_f)(\psi_f,\psi_g)(\psi_g,P_L)(P_L,P_H),
\]
both define cyclic Bargmann structures involving the heavy and light propagation eigenstates together with the decay-projected states. However, the two constructions encode the inter-channel correlations in geometrically distinct ways.

In the disconnected structure described by $\Delta_4$, the decay channels $f$ and $g$ are correlated through successive propagation overlaps involving the heavy and light propagation eigenstates. The resulting invariant therefore characterizes a cyclic inter-channel geometry in which the projected decay states participate through separate overlap chains involving the propagation eigenstates. By contrast, the sequential invariant contains the explicit overlap $(\psi_f,\psi_g)$, introducing a direct geometric correlation between the projected decay states themselves inside the cyclic sequence. The sequential construction therefore probes a connected projection geometry associated with the relative orientation of the decay-induced rays in projective Hilbert space.

The overlap
\begin{align}
(\psi_f,\psi_g)\propto A_f^*A_g+\bar A_f^*\bar A_g
\end{align}
plays an important role in this interpretation. Since this structure depends directly on correlations between the decay projections, the sequential invariant becomes explicitly sensitive to the relative phase alignment between the channels $f$ and $g$. Consequently, while the disconnected invariant primarily probes correlated cyclic interference through the propagation eigenstates, the sequential structure additionally captures information about the relative geometric orientation of the projected decay rays themselves.

These geometric differences are reflected in the behavior of the associated ratio observables near the \CP-conserving regime. The quantity
\begin{align}
R_{seq}
\sim \frac1{|p|^2-|q|^2}
\end{align}
exhibits an enhancement associated with the suppression of the heavy-light overlap factor near exact \CP symmetry. Geometrically, this indicates that the sequential projection overlap continues to influence the correlation structure even as the heavy-light overlap becomes small. The ratio therefore exhibits an inverse enhancement near the limit of vanishing indirect \CP violation within the connected sequential framework.

The comparison ratio $\widetilde R_{seq}=\frac{\Delta_4^{seq}}{\Delta_4}$ provides a complementary measure of the relative behavior of connected and disconnected geometric structures. In the generic small \CP regime,
\begin{align}
    \widetilde R_{seq}\sim |p|^2-|q|^2,
\end{align}
indicating that the relative contribution measured by $\widetilde R_{seq}$ becomes suppressed near \CP symmetry. However, this suppression is modified by the relative interference-phase alignment of the projected decay states, reflecting the interplay between mixing asymmetry and projection coherence. From the projective-space perspective, the transition from the disconnected invariant to the sequential invariant corresponds to replacing an indirect correlation mediated by propagation overlaps with a direct correlation between decay-induced rays.

Taken together, the various Bargmann structures discussed in this work form a geometric hierarchy of \CP-sensitive correlations in neutral meson systems. The third-order invariant characterizes single-channel geometric interference, the disconnected fourth-order invariant describes cyclic inter-channel correlations, while the sequential fourth-order structure captures directly connected projection geometry between decay-induced states. The associated ratios $R_{seq}$ and $\widetilde R_{seq}$ further probe the enhancement properties and comparative behavior of these distinct geometric correlation structures.

\section{Possible Physical Channels and Experimental Relevance}
The geometric structures developed in this work may be relevant for several neutral meson systems in which mixing, interference, and correlated decay processes play an important role \cite{PDG2024, BigiSanda2009}. Since the BIs and the associated ratio observables are constructed from rephasing-invariant overlap structures, the framework is naturally suited for systems where relative interference phases and time-dependent correlations are experimentally accessible. Since the construction relies on decay-projected conditional states arising from correlated meson systems, experimental environments producing entangled neutral meson pairs are particularly natural settings for its application.

A particularly interesting setting for the sequential geometric structures is provided by the neutral kaon system. The $K^0-\overline{K^0}$ sector exhibits small indirect \CP violation together with long-lived interference phenomena, making it an appropriate regime for studying the behavior of the sequential geometric ratios near the \CP-conserving limit \cite{Christenson1964, KLOE2006, BigiSanda2009}. In particular, the enhancement structure $R_{seq} \sim \frac{1}{|p|^2-|q|^2}$ suggests that the associated sequential geometric ratios may exhibit enhanced scaling behavior in regimes where the heavy-light overlap factor becomes small. At the same time, the behavior of the ratio is not governed solely by the mixing asymmetry parameter, since the overlap structures also depend on the relative interference phases and decay-channel correlations appearing in the decay-projected states. Consequently, the sequential geometric structures may provide a complementary geometric characterization of interference in neutral kaon decay processes. Possible realizations include correlated kaon decays involving two-pion, semileptonic, or multi-pion final states, where interference effects between different decay channels are already known to play an important role. In such systems, the connected sequential invariant may be interpreted as probing the geometric relation between decay-induced projection structures within the underlying projective Hilbert space description.

The neutral $B_d^0-\overline{B_d^0}$ system provides another important setting in which the present framework may have phenomenological relevance. Time-dependent \CP asymmetry measurements in $B$-meson decays already probe interference structures associated with mixing and decay phases \cite{BABAR2004, Belle2001, PDG2024}, particularly in channels such as $B^0\to J/\psi K_S$, $B^0\to \pi^+\pi^-$, and related decay modes. In this context, the sequential Bargmann structures introduced here may provide a geometric organization of correlated interference information associated with different decay channels. The sequential invariant depends explicitly on relative phase correlations between the participating decay channels. Consequently, the associated geometric ratios depend not only on mixing asymmetry, but also on the relative alignment of the interference phases appearing in different decay channels. The behavior of both $R_{seq}$ and $\widetilde R_{seq}$ is therefore determined by a combination of mixing-induced \CP violation and phase-dependent overlap coherence.

Similar considerations may also apply to the $B_s^0-\overline{B_s^0}$ system, where rapid oscillations and multi-channel interference structures play an important role in time-dependent analysis \cite{PDG2024}. In particular, channels involving correlated angular or polarization-dependent interference may potentially offer a natural setting for interpreting the sequential Bargmann structures as geometric measures of decay-channel overlap correlations. The framework may therefore offer an alternative geometric characterization of correlated interference structures in neutral meson systems with nontrivial mixing and decay dynamics.

The $D^0-\overline{D^0}$ system may also be of conceptual interest, since both mixing and \CP-violating effects remain small within the Standard Model \cite{PDG2024}. In this regime, the sequential geometric ratios may provide a useful framework for analyzing small deviations from exact symmetry. However, the practical significance of such effects depends on the detailed interference structure of the participating decay channels and cannot be inferred from the mixing asymmetry parameter alone.

Overall, the geometric structures introduced in this work are most naturally connected to time-dependent interference phenomena in correlated neutral meson systems. Rather than defining new observables independent of existing \CP-sensitive quantities, the sequential BIs provide a geometric framework for organizing the underlying interference and phase-correlation structures associated with mixing and decay processes.

\section{Summary}

In this work, we investigated sequential geometric correlation structures in neutral meson systems within the framework of BIs and projective Hilbert space geometry. Building upon previously developed geometric constructions involving propagation eigenstates and decay-projected conditional states, we introduced a connected sequential fourth-order BI in which the projected decay sectors are directly correlated through an explicit projection overlap structure. This construction introduces a new class of connected cyclic overlap structures in which the decay-projected states themselves participate as directly correlated geometric elements.

Unlike the previously studied disconnected fourth-order invariant, where different decay channels are connected only through heavy-light propagation overlaps, the sequential construction considered here incorporates a direct overlap between the projected decay states themselves. This leads to a geometrically distinct cyclic structure associated with connected sequential projection correlations in projective Hilbert space. The resulting invariant therefore provides a complementary geometric description of correlated \CP-sensitive interference structures in neutral meson systems.

To analyze the properties of the sequential geometry, we introduced two rephasing-invariant geometric ratios. The first ratio, $R_{seq}$, characterizes the scaling behavior of the connected sequential structure relative to the third-order invariants associated with the individual decay channels. The second ratio, $\widetilde R_{seq}$, compares the connected sequential invariant directly with the previously constructed disconnected fourth-order structure. Together, these quantities reveal distinct geometric scaling behaviors associated with connected and disconnected overlap structures near the \CP-conserving regime.

A detailed small \CP analysis was performed using the standard rephasing-invariant interference parameters. In this regime, the sequential structures exhibit characteristic scaling behavior governed not only by the mixing asymmetry parameter $|p|^2-|q|^2$, but also by the relative interference phases and geometric alignment of the corresponding decay-projected states. In particular, the sequential invariant depends explicitly on the relative phase alignment between projected decay sectors, reflecting the geometric role of the direct overlap between decay-projected states within the cyclic overlap chain.

The geometric interpretation of the sequential structures was also examined in detail. The previously constructed third-order and disconnected fourth-order BIs were shown to characterize single-channel and cyclic inter-channel geometric correlations, respectively, while the sequential construction introduced here probes directly connected projection geometry between decay-induced states. The resulting framework therefore extends a geometric hierarchy of correlated \CP-sensitive overlap structures in neutral meson systems.

Finally, we discussed possible physical settings in which the present framework may be relevant, including neutral kaon, $B_d$, $B_s$, and $D$ meson systems. Although the present work does not constitute a conventional cascade-decay formalism involving multiple neutral meson propagation sectors, the connected sequential overlap structures studied here provide a complementary geometric viewpoint on correlated interference phenomena and phase-sensitive decay coherence in neutral meson systems.

The results presented in this work suggest that Bargmann invariant methods may provide a useful framework for organizing and interpreting correlated interference structures within a geometric rephasing-invariant formulation. Further investigation of related sequential geometric structures in more general correlated decay systems, including genuinely multi-stage cascade processes and entangled neutral meson sectors, may provide additional insight into the geometric organization of \CP-sensitive quantum interference phenomena.

{\it Acknowledgement :} I wish to thank Prof. Utpal Sarkar and Prof. Arghya Taraphder for support and encouragement. I would like to thank MoE, Government of India for the research fellowship.

\end{document}